\documentclass[12pt]{article}

\usepackage{graphicx}
\graphicspath{{figuras/}}

\usepackage[urlbordercolor={1 1 1}]{hyperref}

\renewcommand{\abstract}[2]{%
      \begin{center} Resumen \end{center} #1\\[1ex]
      \begin{center} Abstract \end{center} {\it #2} 
      \newpage}

\newcommand{\palabra}[1]{\index{#1}#1}
\newcommand{\bfvec}[1]{\mathbf{\vec{#1}}}

\begin{document}

\title{\bf La teor\'ia de supercuerdas}
\author{Herbert Morales\footnote{hmorales@fisica.ucr.ac.cr}\\[1em]
\normalsize \it Escuela de F\'isica\\
\normalsize \it Universidad de Costa Rica\\
\normalsize \it San Jos\'e, Costa Rica
}%
\date{}

\maketitle

\abstract{
Presentamos una breve descripci\'on de la teor\'ia de supercuerdas,
bas\'andonos en la capacidad intuitiva del lector y en su
conocimiento de f\'isica general y c\'alculo a nivel universitario.
Actualmente, la teor\'ia de supercuerdas es ca\-ta\-lo\-ga\-da como
la mejor candidata a explicar los fen\'omenos f\'isicos que
involucren todas las interacciones conocidas:
la gravitacional, la electromagn\'etica, y las nucleares d\'ebil y fuerte.
Nuestra descripci\'on ilustra y explica ciertos conceptos y detalles de
las supercuerdas, tales como supersimetr\'ia, com\-pac\-ti\-fi\-ca\-ci\'on,
dualidades y branas, entre otros.
}{
We discuss superstring theory
based on the reader's intuition and her college
knowledge of general physics and calculus.
Nowadays, superstring theory is considered the
best candidate to explain physical phenomena
that involve all of the known fundamental interactions:
the gravitational, electromagnetic, weak and strong forces.
The main purpose of this work is to show and explain
some of the concepts and details about superstrings,
such as supersymmetry, compactification, dualities
and branes.
}


\section{Introducci\'on}

Las {\it part\'iculas} y usualmente los objetos, al estilo
de los cursos de f\'isica ge\-ne\-ral, se modelan a trav\'es de {\it puntos},
por lo que no importa que tanto nos acerquemos a estas ``part\'iculas",
siguen siendo puntos.
Esta idealizaci\'on presenta el problema de {\it divergencias}
(cantidades f\'isicas que tienden a infinito) en las
{\it teor\'ias cu\'anticas de campo}
(QFT, por sus siglas en ingl\'es).
Las \palabra{QFT} son los modelos para estudiar a las part\'iculas elementales,
como el electr\'on o el fot\'on
(la part\'icula que constituye la luz).
Sin embargo, los m\'etodos de {\it regularizaci\'on y renormalizaci\'on}
en QFT permiten el control de tales divergencias de una forma
sistem\'atica.
Por ejemplo, podemos formular de ma\-ne\-ra consistente y
con poder predictivo
{\it la teor\'ia de electrodin\'amica cu\'antica}
(\palabra{QED}, explica la interacci\'on del fot\'on con las dem\'as part\'iculas), 
{\it la teor\'ia electrod\'ebil}
(o modelo de Glashow-Weinberg-Salam, explica la desintegraci\'on beta) y
{\it la teor\'ia de cromodin\'amica cu\'antica}
(\palabra{QCD}, explica como los protones y los neutrones se mantienen unidos
dentro del n\'ucleo de un \'atomo).
Cabe mencionar que estas teor\'ias juntas conforman el llamado
{\it \palabra{modelo est\'andar}} (SM) de las part\'iculas.

\begin{figure}[!htp]
\begin{center}
\includegraphics[height=1.4in]{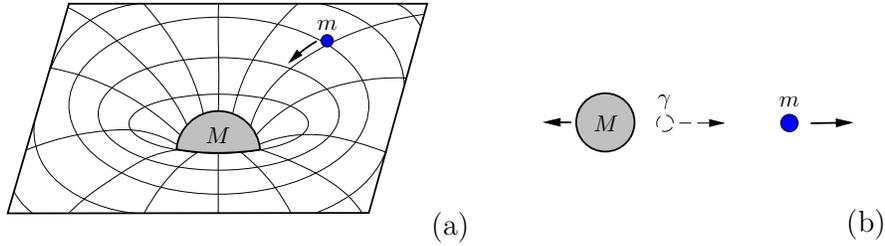}
\end{center}
\caption[La interacci\'on entre dos part\'iculas seg\'un
relatividad general y teor\'ia cu\'antica de campo.]{La
interacci\'on entre dos part\'iculas seg\'un:
(a) la teor\'ia de la re\-la\-ti\-vi\-dad general y (b) la teor\'ia cu\'antica de campo.
En (a) la part\'icula $M$ ``curva" el espacio y atrae a $m$,
mientras en (b) $M$ emite a $\gamma$ (mediadora) y repela a $m$,
cuando esta la absorbe.}
\label{fig:GR-QFT}
\end{figure}

Por otro lado, \index{relatividad general} {\it la teor\'ia general de la relatividad} 
(\palabra{GR}, explica los fe\-n\'o\-me\-nos f\'isicos que involucran la gravedad)
es de una naturaleza completamente diferente que las QFT,
puesto que la interacci\'on gravitacional se modela por medio de un
espacio curvo, mientras las interacciones en \palabra{QFT}
(electromagn\'etica, y nucleares d\'ebil y fuerte)
se modelan a trav\'es de intercambio de \palabra{part\'iculas mediadoras}
(ver Fig.~\ref{fig:GR-QFT}).
Esta diferencia entre las formulaciones de estas teor\'ias hace que
exista una {\it incompatibilidad} entre ellas, es decir,
si estudiamos un fen\'omeno f\'isico que las involucre a ambas,
como los agujeros negros, obtendr\'iamos resultados contradictorios.

En la b\'usqueda de un marco te\'orico unificado de las
QFT con GR, aparecen las {\it \palabra{teor\'ias de cuerdas}} (ST).
Actualmente, consideramos a ST como la candidata m\'as fuerte a una 
{\it teor\'ia cu\'antica de la gravedad}
(nombre gen\'erico de las teor\'ias que unen QFT y GR)
debido a que su formulaci\'on es co\-he\-ren\-te matem\'aticamente
(sin incompatibilidad).
Incluso algunos f\'isicos consideran que las ST logran
posiblemente el objetivo final: la {\it teor\'ia del todo} (TOE).
Sin embargo, existen otras posibilidades no tan populares como 
la {\it gravitaci\'on cu\'antica} (loop quantum gravity) y
otras propuestas que involucran geometr\'ias cu\'anticas
como la {\it geometr\'ia no conmutativa} o la {\it f\'isica difusa} (fuzzy physics).


\section{Una breve historia de las ST}

Alrededor de 1960, hab\'iamos hallado en los aceleradores una gran
variedad de part\'iculas ``elementales", la materia no solo estaba
compuesta de electrones, neutrones y protones.
Sobre todo hab\'iamos encontrado much\'isimo m\'as
\index{hadrones} {\it part\'iculas hadr\'onicas}
(las que experimentan interacci\'on nuclear fuerte, como el
neutr\'on o el prot\'on)
que \index{leptones} {\it lept\'onicas}
(las que no ``sienten" esta in\-te\-ra\-cci\'on, como el electr\'on).
En 1963, Gell-Mann y Zweig proponen un modelo donde clasifican a
estas part\'iculas hadr\'onicas a trav\'es de sus constituyentes,
los {\it \palabra{quarks}}.
La idea del modelo es muy similar a la tabla peri\'odica de los
elementos, ordenamos las part\'iculas de acuerdo a los quarks que
contengan.
Para la d\'ecada de 1970, hab\'iamos desarrollado \palabra{QCD},
la teor\'ia actualmente aceptada para explicar matem\'aticamente las
interacciones entre las part\'iculas hadr\'onicas.

Otro alternativa para explicar dichas interacciones fue la propuesta
por Veneziano, el llamado {\it modelo dual de resonancia}.
Pero este modelo no tuvo \'exito, en general, porque no era renormalizable
(no hay un m\'etodo para controlar las divergencias)
y porque presentaba {\it anomal\'ias}
(simetr\'ias o leyes de conservaci\'on, como la de momento lineal,
que se cumplen a nivel cl\'asico, pero no al cu\'antico).
Al estudiar este modelo, encontrar\'iamos que sus
{\it ecuaciones de movimiento}
describ\'ian entes extendidos de una dimensi\'on
({\it cuerdas}, en lugar de puntos)
y que predec\'ia la existencia del {\it \palabra{gravit\'on}}
(part\'icula mediadora de la gravedad, cuya presencia no es deseable
para una teor\'ia puramente hadr\'onica).
Este \'ultimo resultado hizo que la interpret\'aramos como una teor\'ia
cu\'antica de la gravedad en lugar de una hadr\'onica,\footnote{%
En los \'ultimos a\~nos, la investigaci\'on de cuerdas como
teor\'ia hadr\'onica se ha reactivado, pero todav\'ia no existe una
formulaci\'on concreta y  plausible.}
desarroll\'andose as\'i la llamada {\it teor\'ia de cuerdas bos\'onicas}.
Sin embargo, los problemas anteriores de renormalizaci\'on y de
anomal\'ias hicieron que esa iniciativa fuera abandonada.

A partir de 1984, logramos formular la
{\it versi\'on supersim\'etrica de la teor\'ia de cuerdas},
o simplemente {\it supercuerdas}, con el resultado que no es \'unica,
pues determinamos {\it cinco teor\'ias} consistentes, sin
contradicciones l\'ogicas y libres de anomal\'ias.
En f\'isica, pensamos que al final tendremos una \'unica teor\'ia que
explique todo lo que est\'e en el universo, de ah\'i que hubo cierta
desaz\'on debido a la existencia de estas cinco teor\'ias.
A este periodo lo conocemos como la
{\it primera revoluci\'on de las cuerdas}.

A partir de 1994, encontramos que las diferentes versiones de las
supercuerdas (teor\'ias de diez dimensiones)
las podemos {\it unificar} a trav\'es de equivalencias
(existen identificaciones uno a uno de los  componentes de dos teor\'ias
supuestamente dispares, ver secci\'on \ref{sec:dualidades})
y que est\'an conectadas dentro de una teor\'ia {\it madre} de once
dimensiones, llamada la {\it \palabra{teor\'ia M}}.
La idea de teor\'ia \'unica vuelve de alg\'un modo;
aunque partamos de formulaciones matem\'aticas diferentes,
pero ahora relacionadas por las equivalencias.
A este periodo lo conocemos como la
{\it segunda revoluci\'on de las cuerdas}.
Entre los nuevos ingredientes que trae la unificaci\'on est\'an
las {\it \palabra{branas}}
(entes f\'isicos de $p$-dimensiones, ver secci\'on \ref{sec:branas})
que permiten nuevos resultados como la
{\it co\-rres\-pon\-den\-cia AdS/CFT}
(una realizaci\'on concreta del principio hologr\'afico,
podemos estudiar un volumen del espacio a trav\'es de su frontera)
y el entendimiento microsc\'opico de la
{\it termodin\'amica de los agujeros negros}.


\section{Part\'iculas puntuales cl\'asicas}

Como mencionamos anteriormente, las part\'iculas las modelamos
normalmente a trav\'es de puntos.
As\'i, si dibujamos la trayectoria descrita por el movimiento de
la part\'icula puntual en el espacio-tiempo, tenemos una curva,
llamada {\it \palabra{l\'inea de mundo}}.

\begin{figure}[!htp]
\begin{center}
\includegraphics{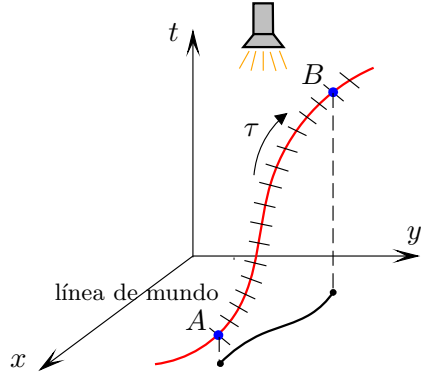}
\end{center}
\caption{Trayectoria de un punto en el espacio-tiempo.}
\label{fig:linea}
\end{figure}

La Fig.~\ref{fig:linea} muestra la l\'inea de mundo para una
part\'icula que se mueve en dos dimensiones (en el plano $x$-$y$).
Note que la coordenada vertical $t$ denota el tiempo,
de ah\'i que el sistema de coordenadas representa el espacio-tiempo.
Por tanto, la l\'inea de mundo no representa el camino por donde pas\'o
la part\'icula, este camino es su proyecci\'on en el plano $x$-$y$
(la sombra que deja la l\'inea de mundo, si la ``alumbramos" desde arriba).
Por ejemplo, el punto $A$ podr\'ia representar su casa y
el punto $B$, su lugar de trabajo; de modo que $A$ estar\'ia dado por
la latitud y longitud donde encontrar\'iamos su casa y por la hora que
usted sale hacia su trabajo, mientras $B$, por la latitud y longitud donde
encontrar\'iamos su oficina y por la hora que usted llega a trabajar.
El largo de la curva que est\'a en el plano $x$-$y$ ser\'ia la distancia
que usted recorre de su casa a su oficina.

Con esta figura, podemos ver que los puntos no son convenientes para
representar part\'iculas u objetos, puesto que matem\'aticamente no
tienen dimensiones, lo que har\'ia posible ``poner" dos part\'iculas en
un mismo punto del espacio-tiempo, es decir,
dos part\'iculas podr\'ian estar en el mismo lugar al mismo tiempo,
lo que no es f\'isicamente posible.
Esta situaci\'on de estar en el mismo punto del espacio-tiempo aparece
como divergencias en las \palabra{QFT}.
Aunque existen m\'etodos para controlar estas divergencias, una 
formulaci\'on m\'as adecuada de la teor\'ia deber\'ia incluir alguna
dimensi\'on a las part\'iculas.

\begin{figure}[!htp]
\begin{center}
\includegraphics{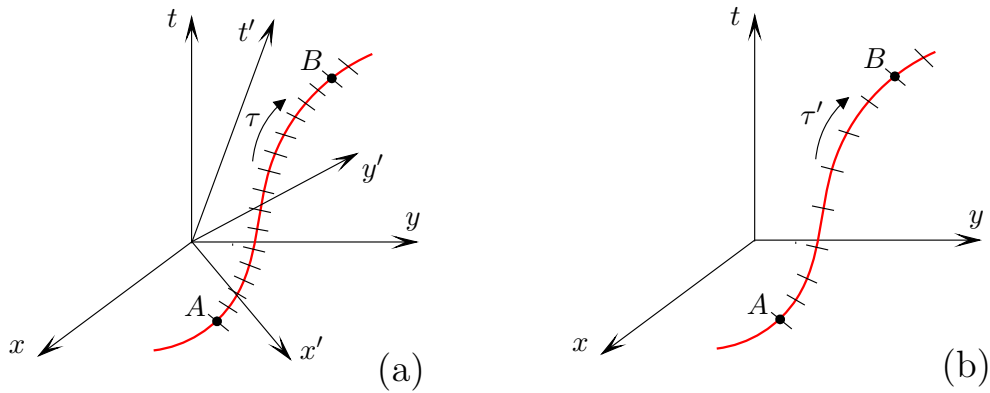}
\end{center}
\caption[Las simetr\'ias de las part\'iculas puntuales.]{Las
simetr\'ias de las part\'iculas puntuales:
(a) la de Lorentz y (b) la de parametrizaci\'on.
Compare con la Fig.~\ref{fig:linea}.}
\label{fig:sim-pts}
\end{figure}

Cabe mencionar que nuestra representaci\'on puntual de las
part\'iculas tiene las siguientes \index{simetr\'ia} {\it simetr\'ias}
(transformaciones o cambios que le podemos hacer al sistema de
coordenadas o a los objetos, de manera que no alteremos la
interpretaci\'on del fen\'omeno f\'isico):

\begin{itemize}

{\it \item simetr\'ia del espacio-tiempo}
[Fig.~\ref{fig:sim-pts}(a)]:
la escogencia del sistema de coordenadas es arbitraria, as\'i que
podr\'iamos rotar este sistema
(incluso hacer que los ejes no sean ortogonales)
sin consecuencia alguna sobre la l\'inea de mundo.
Tambi\'en, es conocida como {\it simetr\'ia de Lorentz}.
Si adem\'as permitimos que el origen del sistema de coordenadas se
ubique en otro lugar, la llamar\'iamos {\it de Poincar\'e}.

{\it \item simetr\'ia de parametrizaci\'on de la l\'inea de mundo}
[Fig.~\ref{fig:sim-pts}(b)]:
el modo c\'omo subdividimos esta l\'inea es arbitrario, por lo tanto,
cualquier subdivisi\'on escogida no debe cambiar la forma de esta l\'inea.
Estas subdivisiones las escogemos preferiblemente para representar la
unidad  de tiempo que ``siente" la part\'icula, el llamado
{\it tiempo propio} $\tau$.

\end{itemize}

El concepto de \palabra{simetr\'ia} (o leyes de conservaci\'on)
es muy importante para la formulaci\'on
de teor\'ias, pues los f\'isicos te\'oricos las usan como puntos de
partida, incluso la mayor\'ia de ellos creen que ciertas simetr\'ias
seguir\'an siendo v\'alidas dentro de la teor\'ia del todo.


\section{\index{cuerdas bos\'onicas}Cuerdas bos\'onicas cl\'asicas \label{sec:string}}

Cambiemos nuestro modo de modelar las part\'iculas, usemos cuerdas
en lugar de puntos.
Ahora si dibujamos la trayectoria descrita por el movimiento de la 
part\'icula en el espacio-tiempo, tenemos una banda,
llamada {\it \palabra{hoja de mundo}}.

\begin{figure}[!htp]
\begin{center}
\includegraphics{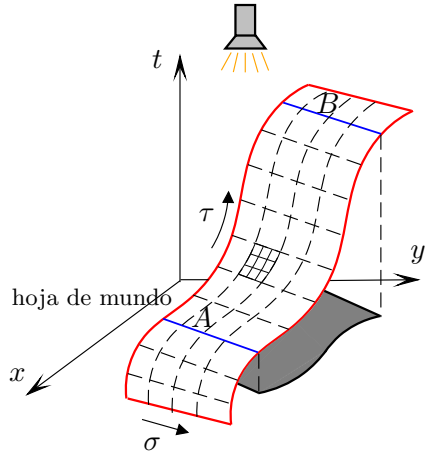}
\end{center}
\caption{Trayectoria de una cuerda en el espacio-tiempo.}
\label{fig:hoja}
\end{figure}

La Fig.~\ref{fig:hoja} muestra la hoja de mundo para una part\'icula
que se mueve en dos dimensiones (en el plano $x$-$y$).
An\'alogamente a la l\'inea de mundo, la hoja de mundo no representa
el camino por donde pas\'o la part\'icula, este camino (banda)
es su proyecci\'on en el plano $x$-$y$
(la sombra de la hoja de mundo alumbrada desde arriba).
Repitiendo el ejemplo anterior, la l\'inea $A$ podr\'ia representar su
casa y la l\'inea $B$ su lugar de trabajo
(ahora si le damos dimensiones a su persona a trav\'es de su ancho,
la distancia entre sus hombros).
La distancia que usted recorre de su casa a su oficina estar\'ia dada
por el largo de la banda que est\'a en el plano $x$-$y$.
Obviamente, el modelaje apropiado de objetos macrosc\'opicos
requerir\'ia de tres dimensiones
(no una, como lo es una cuerda),
pero como nos preocuparemos de las part\'iculas elementales podremos
proponer que son de una dimensi\'on;
al final al ser elementales no pueden estar formadas de algo m\'as
(por eso es elemental),
incluso los experimentos no pueden ``ver" la forma de estas part\'iculas,
as\'i que ``ser\'an" cuerdas en el sentido de que nuestra teor\'ia
reproduzca exactamente los datos experimentales.
Bajo esta premisa, las cuerdas tendr\'an un largo caracter\'istico
(extremadamente peque\~no, llamado la escala de Planck)
que limita completamente los experimentos, no podremos medir
distancias m\'as peque\~nas que ese largo.
Al final, la consistencia matem\'atica de la teor\'ia nos obligar\'a
a incluir entes de dimensiones mayores que uno
(las branas, ver secci\'on \ref{sec:branas}),
as\'i que habr\'an part\'iculas asociadas a estos entes.

Veamos las ventajas de usar una cuerda en lugar de un punto como
re\-pre\-sen\-ta\-ci\'on de una part\'icula en la Fig.~\ref{fig:hoja}:

\begin{enumerate}

\item La hoja de mundo requiere de dos par\'ametros $\tau$ y
$\sigma$ para ser descrita en el espacio-tiempo, recuerde su
curso de c\'alculo las superficies las parametrizamos con dos variables,
mientras las curvas con una variable.
Entonces, el nuevo par\'ametro $\sigma$ podr\'a ser usado para darle
dimensi\'on (tama\~no) a las part\'iculas que era nuestro objetivo
inicial.

\item El cambio de la representaci\'on de las part\'iculas nos hace
involucrar una nueva simetr\'ia, adem\'as de las anteriores.
Ahora tenemos:

\begin{itemize}

{\it \item simetr\'ia del espacio-tiempo o de Lorentz}:
la escogencia del sistema de coordenadas sigue siendo arbitraria.

{\it \item simetr\'ia de parametrizaci\'on de la hoja de mundo}:
el modo c\'omo subdividimos la hoja, tanto a lo ancho como a lo largo,
es arbitrario, as\'i cualquier otra subdivisi\'on en la hoja no cambia
su forma.
En otras palabras, el modo c\'omo cuadriculamos la hoja es arbitrario,
as\'i cualquier otro cuadriculado en la hoja no cambiar\'a su forma,
incluso la cuadr\'icula no tiene que ser de cuadrados,
podr\'ian ser de rombos o cualquier otro cuadril\'atero
[figura geom\'etrica de cuatro lados, ver Fig.~\ref{fig:sim-st}(a)].
Las subdivisiones del largo podr\'an representar
las unidades del tiempo propio $\tau$;
mientras las del ancho, las unidades de longitud propia $\sigma$.

{\it \item simetr\'ia de la m\'etrica local}:
el concepto de m\'etrica es muy importante cuando se trabaja con
coordenadas generalizadas, e.g.~las esf\'ericas,
las rectangulares no son las \'unicas que se pueden usar.
Su uso principal es para determinar apropiadamente
las distancias entre dos puntos o el largo de una curva,
as\'i por ejemplo en coordenadas polares,
escribir\'iamos que el diferencial de arco de un c\'irculo de radio $r$
es $d\ell = r d \theta$ y dir\'iamos que el factor $r$ que nos
permite calcular $d\ell$ a trav\'es de $d\theta$ proviene de la m\'etrica.
Volviendo a la hoja de mundo,
se puede mostrar que podemos reescalar su m\'etrica interna (local)
sin modificar su forma o tama\~no,
ser\'ia como cambiar los patrones de medida internos sin tener
la consecuencia que se cambien los del espacio-tiempo,
en otras palabras dentro de la hoja de mundo no hay escalas
(decir km o m es lo mismo).
Esta peculiaridad no es posible en objetos de dimensiones mayores
como las branas.
El nombre t\'ecnico de esta simetr\'ia es
{\it de Weyl o de reescalamiento local de la m\'etrica}
[ver Fig.~\ref{fig:sim-st}(b)].

\end{itemize}

\begin{figure}[!htp]
\begin{center}
\includegraphics{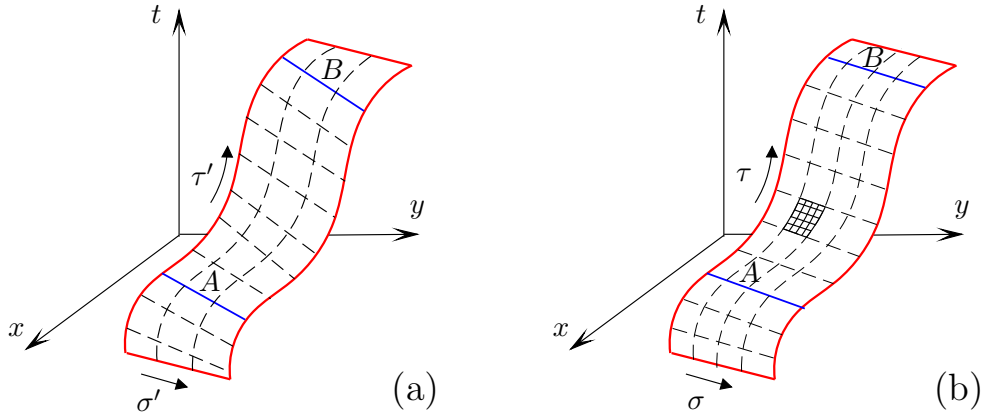}
\end{center}
\caption[Las simetr\'ias de las cuerdas bos\'onicas.]{Las
simetr\'ias de las cuerdas bos\'onicas:
(a) la de parametrizaci\'on y (b) la de Weyl.
Compare con la Fig.~\ref{fig:hoja} y
note que la primera simetr\'ia modifica el cuadriculado,
mientras la segunda lo deja igual,
pero ahora los cuadros son de $4\times5$,
en lugar de $3\times3$ en unidades vistas dentro de la hoja mundo;
en el espacio-tiempo los cuadros no cambian sus dimensiones.}
\label{fig:sim-st}
\end{figure}

\item A diferencia de los puntos, las cuerdas tienen dimensi\'on,
sus largos; por lo que podr\'iamos ``jugar" con sus puntas
para formar dos \palabra{tipos de cuerdas}:
{\it las abiertas y las cerradas} (ver Fig.~\ref{fig:cuerdas}).
Por tanto, habr\'ia dos ``tipos" de part\'iculas de acuerdo a si las
modelamos como cuerdas abiertas
(ej.~{\it el fot\'on}, la part\'icula que  media la interacci\'on
electromagn\'etica)
o cerradas
(ej.~{\it el \palabra{gravit\'on}}, la que media la gravitacional).
\begin{figure}[!htp]
\begin{center}
\includegraphics{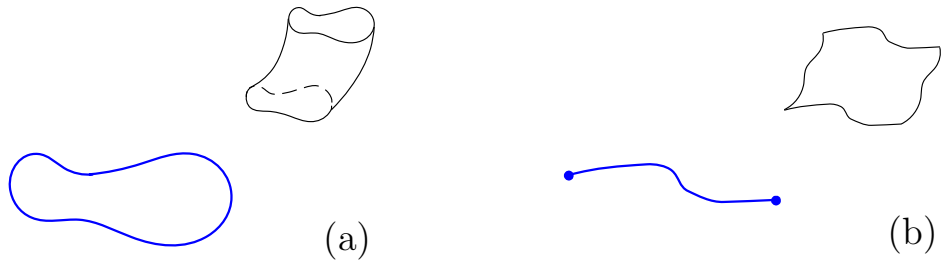}
\end{center}
\caption[Clasificaci\'on de la cuerda.]{Clasificaci\'on
de la cuerda de acuerdo a si est\'a:
(a) cerrada o (b) abierta.
Al lado la hoja de mundo respectiva.}
\label{fig:cuerdas}
\end{figure}

\item Podr\'iamos tambi\'en hacer vibrar a la cuerda, en el
sentido que haya una onda (o varias) traslad\'andose en ella, como en
una cuerda de guitarra o cuando uno toma un extremo de un mecate que
est\'a atado en su otro extremo y lo comienza a oscilar.
El objetivo te\'orico es que cada {\it modo de vibraci\'on} sirva para
definir diferentes part\'iculas. \index{tipos de cuerdas}
Por ejemplo, el modo de vibraci\'on {\it uno} podr\'ia definir a un
electr\'on y el modo {\it dos}, a un mu\'on
(part\'icula similar al electr\'on, pero con mayor masa).
En cuerdas cerradas, clasificamos estas vibraciones de acuerdo a la
direcci\'on en que se propagan las ondas, las llamaremos {\it derechas}
a las que van en direcci\'on contraria a las manecillas de reloj e
{\it izquierdas} a las que van a favor, vistas de frente al movimiento
de la cuerda
(ver Fig.~\ref{fig:mano}).
Para cuerdas abiertas, esta clasificaci\'on no tiene sentido porque
sus puntas ``devuelven" o ``rebotan" a las ondas, por tanto,
al llegar a una de las puntas la direcci\'on de la
propagaci\'on cambia, estrictamente tendremos una superposici\'on de
ambas ondas (una onda estacionaria).
\begin{figure}[!htp]
\begin{center}
\includegraphics{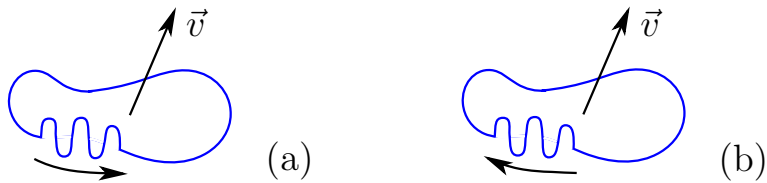}
\end{center}
\caption[Las ``manos" de las cuerdas cerradas.]{Las
``manos" de las cuerdas cerradas:
(a) las derechas y (b) las izquierdas.
Note que si su pulgar apunta en la direcci\'on de $\bfvec{v}$ y
el resto de sus dedos en la de la propagaci\'on de las ondas, 
usted podr\'a correlacionar estas cuerdas con sus respectivas manos.}
\label{fig:mano}
\end{figure}

\end{enumerate}

Con todas las ideas anteriores, formulamos la teor\'ia de cuerdas bos\'onicas.
Una escogencia particular de las simetr\'ias nos permite simplificar
las ecuaciones de movimiento como ecuaciones de onda
(las mismas del curso de f\'isica general).
Entonces, decimos que {\it fijamos la calibraci\'on} ({\it gauge-fixing})
cuando exigimos una escogencia particular entre todas las posibles.
Con esta calibraci\'on, obtendr\'iamos la soluci\'on de las ecuaciones
a trav\'es de una {\it expansi\'on de modos normales}
(representaci\'on de una vibraci\'on general como una superposici\'on
de osciladores arm\'onicos simples, al estilo de una serie de Fourier
que representa cierta funci\'on peri\'odica). 
Para cuantizar la teor\'ia
(reescribir esta soluci\'on con operadores),
lo har\'iamos usando las reglas de conmutaci\'on usuales
(como las del curso de mec\'anica cu\'antica)
que determinar\'ian el \'algebra de los operadores de subida y bajada de
la teor\'ia.
Lo que contin\'ua es entonces definir el estado base y con los operadores
de subida cons\-trui\-r\'ia\-mos todos los otros estados f\'isicos,
lo que permite establecer completamente la teor\'ia al nivel cu\'antico.
Finalmente, la consistencia matem\'atica de la teor\'ia requiere que
nuestra soluci\'on completa (operadores, estados, ...)
respete la simetr\'ia de Lorentz, lo que obliga a
introducir {\it dimensiones extras} al espacio-tiempo, ya no son tres
dimensiones de espacio y una de tiempo, sino 25 de espacio y una de
tiempo.
El mecanismo para lidiar con las dimensiones extras es la
\palabra{compactificaci\'on}
(ver secci\'on \ref{sec:compa}) y su interpretaci\'on f\'isica estar\'ia
asociada con la estructura de nuestro universo cuadridimensional y
las part\'iculas que contiene. 

El an\'alisis de c\'omo transformamos, bajo la simetr\'ia de Lorentz,
los estados f\'isicos sin masa
conlleva a identificarlos con las siguientes part\'iculas:
el gravit\'on, el fot\'on, el {\it dilat\'on}
(la que determina la ``intensidad" de aco\-pla\-mien\-to entre las cuerdas)
y una especie de ``fot\'on" tensorial
(asociado a un tensor antisim\'etrico, llamado campo de Kalb-Ramond).
Sin embargo, existe un ``peque\~no problema", la presencia de
{\it \palabra{taquiones}}
(``part\'iculas" que pueden viajar m\'as r\'apido que la luz).
Esta situaci\'on no es deseable f\'isicamente, porque de acuerdo a la
teor\'ia de la relatividad, no existe objeto alguno que viaje m\'as
r\'apido que la luz.
Por tanto, como deseamos una teor\'ia unificadora que incluya
relatividad, esta caracter\'istica hace que la debamos descartar.
Aunque existe una interpretaci\'on diferente de estos estados
taqui\'onicos que permite eludir este problema y retomar la teor\'ia
f\'isicamente
(la llamada {\it condensaci\'on taqui\'onica}, donde considerar\'iamos
al estado del vac\'io de la teor\'ia como inestable debido a un
rompimiento espont\'aneo de una de sus simetr\'ias).
Nuestra formulaci\'on anterior de las cuerdas bos\'onicas no ser\'ia
entonces la adecuada y requerir\'ia de una reformulaci\'on
que involucrar\'ia tambi\'en branas desde el principio.

\begin{figure}[!htp]
\begin{center}
\includegraphics{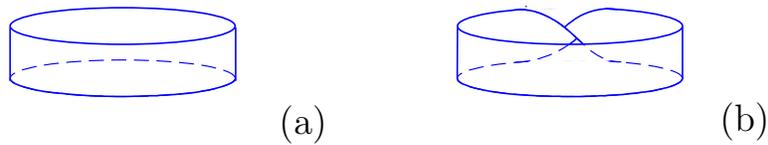}
\end{center}
\caption[Clasificaci\'on de las cuerdas cerradas.]{Clasificaci\'on
de las cuerdas cerradas de acuerdo a si su hoja de mundo:
(a) est\'a orientada o (b) no est\'a orientada.
El caso (b) es cuando uno toma una tira de papel y le da una media vuelta
antes de pegarle sus bordes, en matem\'atica la llamamos la
\it{banda de M\"obius}.}
\label{fig:orienta-cs}
\end{figure}

\begin{figure}[!htp]
\begin{center}
\includegraphics{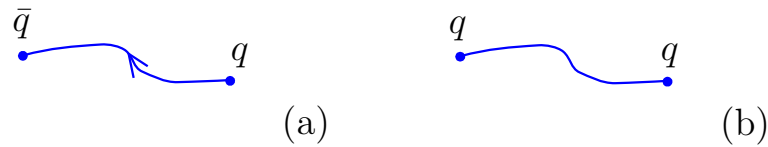}
\end{center}
\caption[Clasificaci\'on de las cuerdas abiertas.]{Clasificaci\'on
de las cuerdas abiertas de acuerdo a:
(a) si est\'an orientadas o (b) no lo est\'an.
$\bar{q}$ denota la carga opuesta a $q$.}
\label{fig:orienta-op}
\end{figure}

Las cuerdas tienen {\it \palabra{orientaci\'on}}, \index{tipos de cuerdas}
la cual podemos visualizar de la siguiente manera.
En el caso de cuerdas cerradas
(ver Fig.~\ref{fig:orienta-cs}),
uno imagina que coloca una hormiga en la parte externa de la
\palabra{hoja de mundo} y esta comienza a caminar,
entonces si al completar una vuelta a la hoja,
la hormiga regresa al punto de partida,
implicar\'a que la cuerda est\'a orientada;
en caso contrario, no est\'a orientada,
la hormiga necesita dar dos vueltas para regresar al punto de partida.
Para cuerdas abiertas, definimos la orientaci\'on a trav\'es de
sus puntas a las que les asignamos cargas
(no tener, significa que la carga es cero).
Estas \index{carga} cargas no son las el\'ectricas
(denominadas como factores de Chan-Paton), 
pero tienen propiedades similares
y adem\'as permiten incorporar otras simetr\'ias a la teor\'ia,
por ejemplo, la simetr\'ia de color, la que determina c\'omo
interact\'uan los \palabra{quarks} entre s\'i.
Entonces, si las cargas son iguales,
diremos que la cuerda no est\'a orientada 
(no podemos distinguir sus puntas)
y si son diferentes, que s\'i lo est\'a
(ver Fig.~\ref{fig:orienta-op}).

Por tanto, podemos clasificar las cuerdas seg\'un esta orientaci\'on.
Ma\-te\-m\'a\-ti\-ca\-mente, podemos mostrar que las cuerdas sin orientaci\'on
no pueden interactuar con las orientadas y viceversa, la interacci\'on
entre las cuerdas requieren que sean de una misma ``especie" dentro
de la formulaci\'on de una teor\'ia.
Por ejemplo, en las cuerdas cerradas no existe ninguna manipulaci\'on
sobre la hoja
(una transformaci\'on continua, como estirarla o doblarla)
que permita convertir una hoja orientada en una que no lo es
(el \'unico modo es cort\'andola y peg\'andole los bordes nuevamente,
pero este proceso en la hoja est\'a prohibido porque no es continuo).
Esta prohibici\'on sobre la orientaci\'on m\'as el hecho que las cuerdas
abiertas pueden interactuar y formar cuerdas cerradas
(la manipulaci\'on de unir las cuerdas es siempre permitida, pero no
necesariamente la de romperlas o cortarlas)
hacen que podamos formular cuatro teor\'ias bos\'onicas diferentes:
\begin{enumerate}
\item la de cuerdas cerradas orientadas:
      $G_{\mu \nu},\ B_{\mu \nu},\ \Phi$;
\item la de cuerdas cerradas sin orientaci\'on:
      $G_{\mu \nu},\ \Phi$;
\item la de cuerdas cerradas y abiertas orientadas:
      $G_{\mu \nu},\ B_{\mu \nu},\ \Phi,\ A_\mu$; y
\item la de cuerdas cerradas y abiertas sin orientaci\'on:
      $G_{\mu \nu},\ \Phi$;
\end{enumerate}
donde $G_{\mu \nu}$ denota al \palabra{gravit\'on},
$B_{\mu \nu}$ al ``fot\'on" tensorial,
$\Phi$ al dilat\'on y $A_\mu$ al fot\'on.
Hemos solo indicando cuales part\'iculas sin masa est\'an presentes en
cada teor\'ia.
Note que en todas el gravit\'on est\'a presente, de ah\'i que digamos
que una teor\'ia de cuerdas ``predice" la existencia del gravit\'on
(es una consecuencia) o equivalentemente siempre tenemos a GR.
La presencia del taqui\'on en todas estas teor\'ias sigue siendo el
tal\'on de Aquiles.

\begin{figure}[!htp]
\begin{center}
\includegraphics{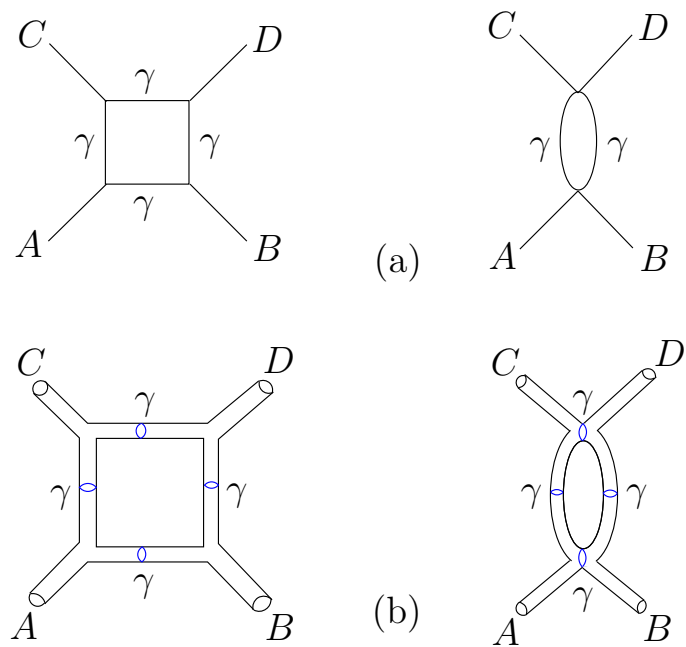}
\end{center}
\caption[Interacci\'on entre dos part\'iculas modeladas como puntos
y cuerdas cerradas.]{Interacci\'on
entre dos part\'iculas modeladas como:
(a) puntos y (b) cuerdas cerradas.
Note que el diagrama de cuerdas cerradas lo obtenemos al ``inflar" el
de QFT.}
\label{fig:interaccion}
\end{figure}

\begin{figure}[!htp]
\begin{center}
\includegraphics[width=5.4in]{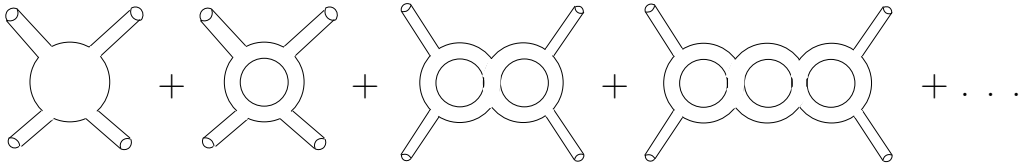}
\end{center}
\caption{Interacci\'on completa entre dos cuerdas cerradas y orientadas.}
\label{fig:doscuerdas}
\end{figure}

Otra consecuencia importante de una teor\'ia de cuerdas es el hecho que
existe una ``\'unica forma" de dibujar (y por tanto, de calcular)
una in\-te\-ra\-cci\'on entre las part\'iculas
(ver Figs.~\ref{fig:interaccion} y \ref{fig:doscuerdas}).
El significado de \'unica podemos entenderlo as\'i:
dado un tipo de interacci\'on entre las cuerdas,
este puede representar diferentes interacciones entre diferentes
part\'iculas, donde la escogencia de los
modos de vibraci\'on determina de cu\'al interacci\'on en particular
estamos hablando.
Esto en contraposici\'on con la representaci\'on puntual que requiere de
la introducci\'on de ciertas funciones espec\'ificas asignadas a cada una
de las part\'iculas, llamadas {\it propagadores} y representadas por
l\'ineas en la Fig.~\ref{fig:interaccion}(a),
el propagador de la $A$ no es necesariamente igual al del $B$.
Desde el punto de vista de cuerdas, el propagador es \'unico para
todas las part\'iculas que la cuerda pueda representar;
contrario a \palabra{QFT}, donde es espec\'ifico a cada una de ellas.
Por ejemplo, la Fig.~\ref{fig:interaccion} muestra el proceso donde
dos part\'iculas incidentes $A$ y $B$ interact\'uan a trav\'es del
intercambio de las \index{part\'iculas mediadoras} mediadoras
$\gamma$ y resultan en $C$ y $D$ (las emergentes).
En el caso (a), las part\'iculas son vistas como puntos, note entonces
que los dos \index{diagramas de Feynman} diagramas (de Feynman)
son diferentes, el de la izquierda
intercambia cuatro $\gamma$ y el de la derecha, solo dos.
Mientras en el caso (b), vistas como cuerdas, los diagramas son los
mismos, la ``desaparici\'on" de los tubos horizontales no cambian
(topol\'ogicamente) los dibujos, el de la derecha los ``tiene",
pero encogidos.
Este es el tipo de argumento que implica una ``\'unica forma".
Incluso note que el punto de uni\'on de las l\'ineas en (a),
llamado v\'ertice, no aparece en (b);
evitando para efectos de c\'alculo un
problema de divergencia al ``poner" los propagadores en un mismo punto
del espacio-tiempo, lo que permitir\'a luego incorporar la interacci\'on
gravitacional sin incoherencias matem\'aticas.

Estas equivalencias (topol\'ogicas) entre los diagramas en una teor\'ia
de cuerdas conlleva tambi\'en a que el n\'umero de los mismos disminuya
en comparaci\'on con los presentes en \palabra{QFT}
(sin importar la orientaci\'on de las cuerdas o si son abiertas,
aunque en estos casos hay m\'as diagramas que con las ce\-rra\-das
orientadas).
La Fig.~\ref{fig:doscuerdas} muestra la interacci\'on completa
entre las dos part\'iculas vistas como cuerdas, cada dibujo es
un \index{diagramas de Feynman} diagrama de Feynman
que involucra las diferentes posibilidades
en que la interacci\'on se puede dar debido a la cantidad de
part\'iculas mediadoras que se intercambian.
As\'i el primer diagrama intercambia una mediadora
(el tubo intermedio lo hemos ``inflado" para que parezca una bola);
el segundo, cuatro mediadoras; el tercero, siete\ldots\
En este caso particular,
la teor\'ia de cuerdas est\'a formulada con la regla de interacci\'on:
solo la uni\'on directa de tres tubos es permitida,
i.e.~la interacci\'on directa involucra a tres cuerdas cerradas y orientadas.


\section{Introduciendo supersimetr\'ia}

Como vimos en la secci\'on anterior, la presencia del taqui\'on hizo
que nuestro plan de una teor\'ia unificadora a trav\'es de cuerdas
bos\'onicas fracasara o al menos no sabemos actualmente c\'omo lidiar
con estos estados taqui\'onicos para formularla adecuadamente.
Por tanto, intentemos poner nuestro esfuerzo en generalizar nuestra
formulaci\'on de teor\'ias de cuerdas con la herramienta de siempre en
f\'isica, simetr\'ias, y con el objetivo de unificar.

En nuestra formulaci\'on puntual de \palabra{QFT}, existe una propiedad f\'isica,
llamada {\it \palabra{esp\'in}}, que permite clasificar las part\'iculas en
{\it \palabra{bosones}} y {\it \palabra{fermiones}}.
Los primeros tienen la peculiaridad de poderse aglomerar entre ellos sin
l\'imite alguno en un volumen dado
(ser\'ian como fantasmas ``llenando" un cuarto)
y los segundos no tienen esa capacidad, s\'i tienen l\'imite
(ser\'ian como personas llenando el cuarto),
a este l\'imite lo conocemos como el {\it principio de Pauli}.
En general, los bosones son las \palabra{part\'iculas mediadoras} de la
in\-te\-ra\-cci\'on, como el fot\'on, y los fermiones son las part\'iculas
materiales, como el electr\'on.
Matem\'aticamente, decimos que los campos que representan
los bosones conmutan (como los n\'umeros reales),
mientras los de los fermiones anticonmutan
(como los vectores en el producto cruz).

Trasladando este concepto a las cuerdas, podemos decir que lo m\'as
ge\-ne\-ral es que hayan unas que sean bos\'onicas y otras, fermi\'onicas
\index{cuerdas fermi\'onicas} 
(de ah\'i el nombre de bos\'onicas \index{cuerdas bos\'onicas} en la secci\'on anterior).
Ahora intentemos ir m\'as all\'a, propongamos una simetr\'ia que nos
diga que para cada fermi\'on debe existir un bos\'on (o viceversa)
de manera que ambos tengan todas las propiedades f\'isicas
(masa, carga el\'ectrica\ldots)
iguales con excepci\'on obviamente del esp\'in.
Esta simetr\'ia de emparejamiento la bautizaremos como
{\it \palabra{supersimetr\'ia}}.
Por ejemplo, la pareja supersim\'etrica del electr\'on es el
{\it s-electr\'on}.
Los nombres de las compa\~neras de las part\'iculas materiales
(fermiones) es simplemente agregando el ``prefijo" {\it s};
en caso de las mediadoras (bosones) es agregando el ``sufijo" {\it ino},
as\'i tenemos la pareja fot\'on-fotino.
Sin embargo, nadie ha visto un s-electr\'on u otra (super)pareja, es
decir, no han sido descubiertos a\'un en los aceleradores de
part\'iculas.
Para explicar porqu\'e no las hemos visto,
introduciremos el concepto de {\it rompimiento de simetr\'ia};
durante la evoluci\'on del universo se dio un proceso que hizo
que la simetr\'ia se perdiera provocando que la masa del
s-electr\'on fuese demasiado grande comparada a la del electr\'on,
por lo tanto, necesitamos energ\'ias muy grandes para
verlo.\footnote{Cabe mencionar que los primeros resultados del LHC,
el m\'as potente acelerador de la actualidad, no han favorecido al concepto de supersimetr\'ia, en otras palabras, las mediciones no han mostrado todav\'ia
la existencia de las superparejas.
Por tanto, si los resultados contin\'uan ``apuntando" a la inexistencia,
la validez de supersimetr\'ia se ver\'a cada vez m\'as limitada al punto
que no tendr\'ia aplicaci\'on alguna a la f\'isica de part\'iculas.}
Por ejemplo, este concepto lo usamos en ferromagnetismo, el hierro es
atra\'ido por imanes si su temperatura es menor que la de Curie
($768^\circ$ C), caso contrario no; la simetr\'ia donde no hay dominios
magn\'eticos se rompe a la temperatura de Curie.

Comencemos entonces a formular una teor\'ia supersim\'etrica de cuerdas,
por simplicidad, consideremos solo cuerdas cerradas
(recuerde de la secci\'on anterior que este tipo de cuerdas debe
estar siempre en la formulaci\'on de la teor\'ia).
Por tanto, dentro de la nueva propuesta tendr\'iamos las
\index{cuerdas bos\'onicas} bos\'onicas,
como las que ya hemos estudiado anteriormente,
m\'as sus superparejas: las \index{cuerdas fermi\'onicas} fermi\'onicas.
Siguiendo nuestra proposici\'on de supersimetr\'ia:
todas las pro\-pie\-da\-des son iguales, excepto el esp\'in.
Consecuentemente, implicar\'ia que todos los criterios anteriores
aplicados a las cuerdas bos\'onicas deben ser
tambi\'en aplicables a las fermi\'onicas, as\'i por ejemplo,
habr\'an unas que son derechas y otras izquierdas.
Sin embargo, habr\'a otra propiedad que no tiene que ser igual y
que no fue mencionada previamente:
la {\it \palabra{periodicidad}} de la \palabra{hoja de mundo}.
Decimos que una funci\'on es peri\'odica cuando repite sus valores
cada vez que la evaluamos a cierto intervalo (el periodo),
como lo es la funci\'on seno que se repite cada $2 \pi$,
pero la llamar\'iamos {\it antiperi\'odica},
si repite con cambio de signo,
como ser\'ia la funci\'on seno si la evaluamos cada $\pi$
(suponiendo ahora que el periodo fuese $\pi$)
pues $\sin (\theta + \pi) = -\sin(\theta)$.
Podr\'iamos entonces definir esta propiedad a las hojas mundo siempre
y cuando respete las simetr\'ias de la teor\'ia, sobre todo la de Lorentz.
En nuestro caso, el rol del periodo lo tiene la longitud propia de la
cuerda $\ell$ (la escala de Planck) y
la variable en la que est\'a definida la periodicidad es $\sigma$
(ver Fig.~\ref{fig:hoja}),
es decir, si evaluamos en $\sigma$ y en $\sigma + \ell$,
podr\'iamos obtener el mismo valor
(diremos que la cuerda es peri\'odica)
o un signo de diferencia
(cuerda antiperi\'odica).
El permitir que $\sigma$ pueda tener todos los valores,
originalmente $0 \le \sigma \le \ell$, es lo mismo que hacemos cuando
trabajamos en coordenadas polares, el \'angulo polar $\theta$ est\'a
estrictamente definido entre $0$ y $2 \pi$, pero puede ser extendido a
todos los n\'umeros reales, si no existe ning\'un problema de
ambig\"uedad matem\'atica.
Volviendo a las cuerdas, tenemos que las bos\'onicas solo pueden ser
peri\'odicas y las fermi\'onicas s\'i pueden tener ambas periodicidades.
El motivo principal de esta diferencia es que las cantidades f\'isicas,
como la energ\'ia, calculadas con las fermi\'onicas siempre las
involucran en parejas, mientras para las bos\'onicas no necesariamente.
Por lo tanto, no es una dificultad incorporar la antiperiodicidad en las
fermi\'onicas debido a que nuestro inter\'es de periodicidad es
sobre las cantidades f\'isicas (lo que medimos en los experimentos)
y no sobre la cuerda (lo que modela las part\'iculas), eso s\'i,
la teor\'ia debe emparejar adecuadamente para que as\'i suceda,
i.e.~dos antiperi\'odicas o dos peri\'odicas.
T\'ecnicamente, no usamos las palabras peri\'odica y antiperi\'odica
en su lugar decimos que est\'an en el \index{sectores}
sector de Ramond (o simplemente R)
y en el de Neveu-Schwarz (o NS), respectivamente.

Ahora debemos agregar estos sectores a la formulaci\'on de la teor\'ia,
sin olvidar las simetr\'ias y las diferentes clasificaciones que 
hicimos en la secci\'on anterior.
El procedimiento a seguir es muy parecido a lo que hicimos cuando
introducimos el concepto de orientaci\'on, debemos investigar las
prohibiciones de pasar de un tipo de cuerda a otro.
Analizando la simetr\'ia de ``manos" de las cuerdas cerradas
(ver Fig.~\ref{fig:mano}),
tenemos que si hay izquierdas, deben haber tambi\'en derechas,
por tanto, habr\'an cuatro sectores en relaci\'on a cu\'al sector 
est\'a cada una, lo que respectivamente denominaremos
como NS-NS, NS-R, \mbox{R-NS}, R-R (para las fermi\'onicas).
Para las abiertas, esta simetr\'ia est\'a ausente, por lo que solo
habr\'a dos sectores: R y NS.
Finalmente, introducimos un operador, llamado {\it n\'umero fermi\'onico},
que ``cuenta" por separado cuantas cuerdas derechas o izquierdas hay
en un estado f\'isico, lo que permite definir los sectores `$+$' y `$-$'
en referencia a si hay un n\'umero par o impar de ellas, respectivamente. 
Incluyendo esta subdivisi\'on de las fermi\'onicas, lle\-ga\-mos
a tener estrictamente un total de {\it 16 sectores} diferentes en las
cuerdas cerradas y {\it cuatro} en las abiertas.
Por ejemplo, existen los sectores:
(NS$+$, NS$+$) y (R$-$, NS$+$) en cerradas,
(NS$+$) y (R$-$) en abiertas, entre otros.
En principio podr\'iamos formular un sin n\'umero de teor\'ias si
permitimos combinar a nuestro gusto estos sectores.
Sin embargo, las siguientes ``prohi\-bi\-cio\-nes" (requisitos)
en los \palabra{sectores} restringen enormemente el n\'umero de
teor\'ias de supercuerdas posibles:

\begin{enumerate}

\item Coincidencia en los modos de vibraci\'on entre las derechas
y las izquierdas
({\it level-matching});
la simetr\'ia de mano obliga a la teor\'ia, por ejemplo, a tener tanto
la derecha de modo tres como a la izquierda de modo tres.

\item Los c\'alculos de las interacciones entre los sectores est\'en
bien definidos, sin ninguna ambig\"uedad matem\'atica.

\item El sector resultante de la interacci\'on entre dos sectores debe
incluirse para que la teor\'ia cierre.

\item La teor\'ia debe tener los sectores R-NS y NS-R o en su lugar
el R-R debido a la {\it invariancia modular}, simetr\'ia de los
diagramas de Feynman que representan el estado del vac\'io de la
teor\'ia.

\item Ausencia de taquiones, los sectores que representen esta
part\'icula non grata no deben incluirse en la teor\'ia.

\item La dimensi\'on del espacio-tiempo sea {\it diez} para que la
teor\'ia est\'e libre de anomal\'ias.
\label{req:diez}

\item Que la supersimetr\'ia tambi\'en se refleje en los estados
f\'isicos de la teor\'ia, as\'i al identificarlos con las part\'iculas
establezcamos siempre la pareja supersim\'etrica, por ejemplo,
gravit\'on-gravitino.
Originalmente, la {\it \palabra{supersimetr\'ia} est\'a en las
\index{hoja de mundo} hojas de mundo}
de las cuerdas, as\'i que al ``llevarla" a los estados decimos que
tenemos {\it supersimetr\'ia en el espacio-tiempo}.
Ese proceso de llevarla lo llamamos {\it proyecci\'on GSO}.
\label{req:susy}

\item Prohibici\'on sobre la orientaci\'on, cuerdas orientadas no
interact\'uan con las que no lo est\'an.

\item Cuerdas cerradas siempre estar\'an presentes en cualquier
teor\'ia de cuerdas.

\end{enumerate}

Las \'ultimas dos son las que impusimos en las teor\'ias bos\'onicas.
A pesar de imponer todos estos requisitos, podemos mostrar que
existen cinco opciones viables para formular adecuadamente una
teor\'ia de supercuerdas, las des\-cri\-bi\-re\-mos en la siguiente secci\'on.
Incluso el n\'umero es todav\'ia much\'isimo mayor si consideramos que
la compactificaci\'on no es \'unica, existen varias maneras de lidiar
con las dimensiones extras
(ver secci\'on \ref{sec:compa}).
Aclaramos que en las secciones siguientes, usaremos la palabra
supercuerdas para referirnos tanto a las teor\'ias formuladas con
supersimetr\'ia
(cumplen con el requisito \ref{req:susy})
como a las cuerdas presentes en dichas teor\'ias, el significado
lo seguiremos por contexto.
Si eliminamos el requisito \ref{req:susy}, es posible todav\'ia formular
teor\'ias consistentes, no las llamar\'iamos supercuerdas de acuerdo a
nuestra aclaraci\'on previa, pero las descartamos porque suponemos
que las part\'iculas vienen en superparejas, a pesar de que 
todav\'ia no hemos visto ninguna.


\section{Teor\'ias de supercuerdas}

Como mencionamos en la secci\'on anterior, la supersimetr\'ia permite
formular las supercuerdas en cinco diferentes posibilidades con
consistencia matem\'atica
(de acuerdo a los requisitos previos).
A continuaci\'on describiremos brevemente cada una de ellas.
\index{tipos de teor\'ias de supercuerdas}
\begin{description}

\item[Supercuerdas tipo II-A:] {\ }

Las formulamos con {\it cuerdas cerradas y orientadas}, donde tanto las
cuerdas derechas como las izquierdas tienen por separado su propia
supersimetr\'ia de espacio-tiempo.
La proyecci\'on GSO para esta teor\'ia involucra los sectores NS$+$ y
R$+$ de las derechas contra los sectores NS$+$ y R$-$ de las izquierdas
(decimos que los estados no son quirales).

Si incluimos branas
(ver secci\'on \ref{sec:branas})
dentro de su formulaci\'on, la teor\'ia tambi\'en contiene
{\it cuerdas abiertas y orientadas}.
Con la presencia de las abiertas, podemos incorporar m\'as simetr\'ias
a estas teor\'ias a trav\'es de sus puntas, ver Fig.~\ref{fig:orienta-op}(a).
El requisito \ref{req:diez} de la secci\'on anterior es suficiente para
no tener anomal\'ias.

\item[Supercuerdas tipo II-B:] {\ }

Tiene la misma descripci\'on anterior, sin embargo, la proyecci\'on GSO
involucra solo los sectores con `$+$'
(decimos que los estados son quirales).

\item[Supercuerdas tipo I:] {\ }

Las formulamos con
{\it cuerdas abiertas y cerradas que no est\'an orientadas},
es la versi\'on sin orientaci\'on de las supercuerdas tipo II-B.
La presencia de las abiertas, debido a sus condiciones de frontera,
obliga a que las cuerdas derechas y las izquierdas tengan
la misma supersimetr\'ia de espacio-tiempo.
Estas modificaciones respecto a las de tipo II-B hacen que la teor\'ia
tenga anomal\'ias,
el requisito \ref{req:diez} de la secci\'on anterior no es suficiente,
por lo que requeriremos imponer una simetr\'ia extra, llamada $SO(32)$,
en las puntas de las abiertas para restaurar la consistencia matem\'atica.
Es la \'unica teor\'ia de supercuerdas en donde las cerradas pueden
romperse para originar cuerdas abiertas.
Contiene obligatoriamente branas por consistencia,
debido a que tiene cuerdas abiertas (ver secci\'on \ref{sec:cuerdabrana}).

\item[Supercuerdas heter\'oticas de $SO(32)$:] {\ }

Las formulamos con {\it cuerdas cerradas y orientadas}, donde las
derechas son como las del tipo II
(note que sus sectores tanto en II-A como en II-B son los mismos)
y las izquierdas son solo cuerdas bos\'onicas
(las de la secci\'on \ref{sec:string}),
una heterogeneidad entre teor\'ias de cuerdas.
Por tanto, la supersimetr\'ia de espacio-tiempo solo act\'ua en las
cuerdas derechas, no hay en las izquierdas por no haber cuerdas
fermi\'onicas. 
Para poder ``manejar" las 26 dimensiones de las cuerdas bos\'onicas,
de tal manera que todo ``exista" en las diez dimensiones que implica la
supersimetr\'ia, es necesario imponer la simetr\'ia $SO(32)$
en las cuerdas izquierdas, bajo una modificaci\'on te\'orica apropiada
en ellas para restablecer adecuadamente los requisitos de la secci\'on
anterior debido a que la heterogeneidad no cumple con algunos de ellos.

\item[Supercuerdas heter\'oticas de $E_8 \times E_8$:] {\ }

El manejo de las 26 dimensiones es tambi\'en posible con otra simetr\'ia,
la llamada $E_8 \times E_8$.
De ah\'i que existan dos tipos de teor\'ias heter\'oticas de acuerdo a
la simetr\'ia empleada.

\end{description}


\section{\index{compactificaci\'on} Compactificaci\'on \label{sec:compa}}

Para lidiar con las dimensiones extras, decimos que est\'an
{\it compactificadas},
el tama\~no de estas dimensiones es finito,
como por ejemplo, un tubo infinitamente largo que ser\'ia un
espacio de dos dimensiones
(no consideremos su grueso, por tanto, es una superficie),
donde una de ellas ser\'ia infinita (su largo) y la otra, finita
(su circunferencia, la que est\'a compactificada).\footnote{La idea
de {\it compactificar} fue introducida originalmente por
Kaluza y Klein en un intento por unificar electromagnetismo con GR
durante la d\'ecada de 1920.}
Para efectos f\'isicos, las dimensiones compactificadas son adem\'as
extremadamente peque\~nas para que los experimentos actuales no las
hayan ``visto", en el ejemplo la circunferencia ser\'ia bien peque\~na y
nuestro tubo lo ver\'iamos pr\'acticamente como una l\'inea,
``perdi\'endose" lo hueco.

Basado en las supercuerdas, el universo tendr\'ia diez
dimensiones o 9+1
(notaci\'on que indica m\'as expl\'icitamente el hecho que son nueve
dimensiones espaciales y una para el tiempo),
que denotamos matem\'aticamente como $M^{9+1}$.
Como nuestro universo observable es de cuatro dimensiones o 3+1,
tenemos que restringir las otras seis dimensiones presentes en
las supercuerdas a un espacio compactificado, por lo que denotaremos
como $M^{3+1}$ y $X^6$, respectivamente.
Por tanto, nuestra propuesta la escribimos como:
$ M^{9+1} = M^{3+1} \times X^6 $.
Por relatividad general sabemos que el espacio-tiempo es
din\'amico
(evoluciona, no ha sido siempre el mismo),
as\'i consideramos tambi\'en que al inicio nuestro universo estaba
completamente compactificado y luego se dio un proceso que lo
``descompactific\'o", el {\it Big Bang},
seg\'un las supercuerdas solo debi\'o afectar a
tres de las dimensiones y no a todas.

\begin{figure}[!htp]
\begin{center}
\includegraphics{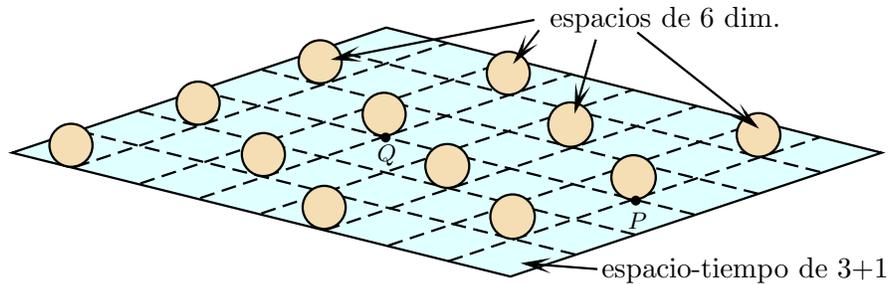}
\end{center}
\caption{Representaci\'on visual de un espacio-tiempo de 9+1.}
\label{fig:fibrado}
\end{figure}
La Fig.~\ref{fig:fibrado} muestra una representaci\'on visual de
este espacio-tiempo, el $M^{9+1}$,
la idea es que en cada punto del espacio-tiempo observable, el $M^{3+1}$,
existe un espacio compactificado, el $X^6$;
as\'i si nos encontramos en el punto $P$ de la figura,
``ver\'iamos" que estrictamente nos hallamos en la superficie de
una esfera $S^6$
(el $X^6$ m\'as simple a escoger).
Note que cada punto tiene su propia esfera, por tanto, la de $Q$ es
otra.
Tambi\'en que el concepto de {\it \palabra{esfera}} es m\'as general al que
habitualmente conocemos, en esta notaci\'on la denotar\'iamos
por $S^2$.
Otro ejemplo: el c\'irculo es la esfera $S^1$.
El super\'indice, por tanto, indica la dimensi\'on de la {\it superficie}
de la esfera, otro concepto m\'as general que lo usamos en referencia
a la ``capa externa", estrictamente la esfera $S^1$ es la circunferencia.
Entre otras posibilidades para el espacio $X^6$, adem\'as de la esfera
$S^6$, y que son matem\'aticamente v\'alidos dentro de la formulaci\'on
te\'orica, est\'an los espacios de Calabi-Yau
(incluyen al toro, nombre matem\'atico para la superficie de una
dona en el caso de dos dimensiones),
las orbivariedades (orbifolds) y los espacios proyectivos.
Debido a esta gama de espacios posibles para $X^6$, tenemos que
estrictamente cada teor\'ia de supercuerdas representa varias, pues
cada uno de estos espacios generan diferentes resultados te\'oricos a
trav\'es de los par\'ametros que los describen, por ejemplo, la esfera
$S^6$ solo incorpora un par\'ametro: su radio.
Sin embargo, una revisi\'on m\'as detallada de estos espacios permite
mostrar que existen equi\-va\-len\-cias entre ellos,
la {\it simetr\'ia de espejo} ({\it mirror symmetry}),
lo que reduce en cierto modo la cantidad de teor\'ias de
supercuerdas.

\begin{figure}[!htp]
\begin{center}
\includegraphics{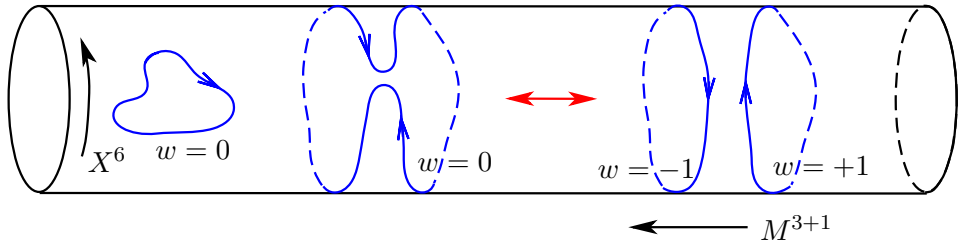}
\end{center}
\caption[Representaci\'on visual de cuerdas cerradas y orientadas con
$w= -1, 0, +1$.]{Representaci\'on
visual de cuerdas cerradas y orientadas con $w= -1, 0, +1$.
La flecha doble indica la transici\'on de una cuerda de $w=0$ a 
dos de $w=-1$ y $w=+1$ (hacia la derecha)
o la transici\'on de dos a una (hacia izquierda).
Las flechas indican el sentido en que se ``enroll\'o", a pesar de que
no se enrolle en la dimensi\'on compactificada. 
La figura estrictamente representa una dimensi\'on compactificada y
la otra no, el tubo infinitamente largo.
Si fuese realmente $X^6$, habr\'ian seis n\'umeros de enrollamiento, uno
por cada dimensi\'on compactificada.}
\label{fig:winding}
\end{figure}

Ahora revisemos si hay alguna consecuencia en las cuerdas debido a la
compactificaci\'on.
Consideremos una dimensi\'on compactificada como en la
Fig.~\ref{fig:winding}, notamos entonces que las cuerdas cerradas
tendr\'ian una nueva propiedad relacionada con el n\'umero de veces
que la cuerda se encuentra enrollada en dicha dimensi\'on;
la bautizaremos como {\it \palabra{n\'umero de enrollamiento}}
({\it winding number}),
la denotaremos con $w$ y le incluiremos signos
para distinguir los dos sentidos de enrollamiento.
Con m\'as cuidado, podemos definir $w$ a las cuerdas abiertas
(ver Fig.~\ref{fig:Dbrana}).
Esta nueva cantidad f\'isica tiene su propia ley de conservaci\'on,
an\'aloga a la del momento lineal, en una interacci\'on entre cuerdas
el n\'umero total de enrollamiento $w_T$ debe permanecer constante
entre el estado inicial y el final.
En la Fig.~\ref{fig:winding}, la transici\'on mostrada tiene $w_T = 0$.
Otra caracter\'istica que adquieren las cuerdas por la compactificaci\'on
es que su \palabra{momento lineal} est\'a ahora cuantizado,
i.e.~sus valores son discretos, al estilo de la carga el\'ectrica
de los cursos de f\'isica general.
En la siguiente secci\'on, veremos c\'omo estas cantidades ayudan a
establecer las dualidades entre las teor\'ias.


\section{Dualidades \label{sec:dualidades}}

La existencia de cinco teor\'ias de supercuerdas desmotiv\'o a los
f\'isicos te\'oricos durante un tiempo, pues no exist\'ia ning\'un
criterio matem\'atico o f\'isico para considerar que una de ellas fuera
la teor\'ia m\'as apropiada en la descripci\'on de la naturaleza.
Fue entonces cuando el descubrimiento de las dualidades reactiv\'o
las supercuerdas, d\'andoles un nuevo aire.

Las \palabra{dualidades} son operaciones matem\'aticas que identifican los
``ingredientes"
(funciones, operadores, estados\ldots)
de dos teor\'ias supuestamente dispares.
Esta identificaci\'on es de uno a uno, cada ingrediente de una de las
teor\'ias tiene su ``pareja" en la otra.
Por ello, decimos que las teor\'ias son {\it equivalentes}:
los sistemas f\'isicos aparentemente diferentes que describen son
estrictamente los mismos.
Como ejemplo considere la teor\'ia electromagn\'etica sin fuentes
(no hay cargas ni corrientes).
La pareja ser\'a el campo el\'ectrico $\bfvec{E}$
y el magn\'etico $\bfvec{B}$.
La dualidad ser\'a entonces que las ecuaciones de Maxwell
(las que definen la teor\'ia)
no cambian bajo la identificaci\'on
$\bfvec{E} \to -\bfvec{B}$ y $\bfvec{B} \to \bfvec{E}$,
i.e.~intercambie $\bfvec{E}$ por $-\bfvec{B}$ y $\bfvec{B}$ por $\bfvec{E}$
en esas ecuaciones y ver\'a que las obtiene de nuevo.
Desempolve su libro de f\'isica y verifique esta ase\-ve\-ra\-ci\'on,
por simplicidad, use la velocidad de la luz igual a uno
($c=1$ o $\epsilon_0 = \mu_0 =1$).
Por tanto, diremos que esta teor\'ia es equivalente (o dual)
a s\'i misma,
la f\'isica (las ecuaciones de Maxwell) sigue siendo la misma
sin importar el intercambio propuesto de los campos.

En algunos casos, establecemos la equivalencia bajo conjetura,
seg\'un un ``emparejamiento" parcial, en otras palabras,
no hay una demostraci\'on formal debido a que no conocemos todos los
ingredientes de las teor\'ias, ver la siguiente secci\'on.
En supercuerdas, clasificamos estas dualidades en tres tipos:
\index{tipos de dualidades}

\begin{description}

\item[Dualidades T] {\ }

Son las que relacionan las teor\'ias de diferentes geometr\'ias
del espacio compactificado sin cambiar las
\index{constante de acoplamiento} {\it constantes de acoplamiento},
i.e.~las teor\'ias presentan la misma
``intensidad en sus interacciones" bajo la dua\-li\-dad.
La m\'as conocida y simple es la que hace la equivalencia de las teor\'ias a
trav\'es del radio de la dimensi\'on compactificada $r$  
(ver Fig.~\ref{fig:Tdual}).
En este caso, establecemos la equivalencia entre una
teor\'ia con radio peque\~no y otra con radio grande.
Particularmente, tenemos la equivalencia entre las teor\'ias de
$R \to \infty$ y de $R \to 0$, una teor\'ia con una dimensi\'on extra
{\it descompactificada} es equivalente a una compactificada con un
radio extremadamente peque\~no.
Consecuentemente, podemos considerar solo teor\'ias con $r \ge \ell$
para cuestiones f\'isicas, al menos en el sentido que nos es m\'as
f\'acil interpretar la f\'isica en estos casos que cuando
$0 \le r \le \ell$, debido a nuestra experiencia en QFT.
De ah\'i, deducimos que la dimensi\'on m\'as peque\~na es $\ell$,
que es consistente con nuestra propuesta de que no podemos ``ver"
m\'as all\'a de la escala de Planck.
En esta dualidad, una pareja de ingredientes es el \palabra{n\'umero de
enrollamiento} y el \palabra{momento lineal} de la cuerda.
Otro ejemplo, similar al de la Fig.~\ref{fig:Tdual}, es entre
las heter\'oticas $E_8 \times E_8 $ y $SO(32)$ y
otros m\'as sofisticados involucran la simetr\'ia de espejo.
\begin{figure}[!htp]
\begin{center}
\includegraphics[width=5.4in]{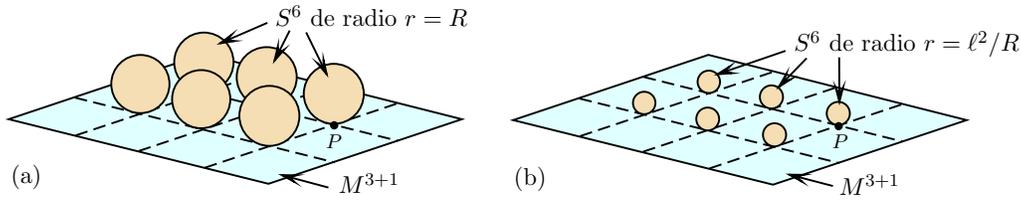}
\end{center}
\caption[La dualidad T entre las teor\'ias tipo II-A y tipo II-B.]{La
dualidad T entre las teor\'ias:
(a) tipo II-A con $r=R$ y (b) tipo II-B con $r=\ell^2/R$,
donde $\ell$ es la escala de Planck.}
\label{fig:Tdual}
\end{figure}

\item[Dualidades S] {\ }

Son las que relacionan una teor\'ia de supercuerdas con
constante de acoplamiento d\'ebil
(cuyos entes f\'isicos dominantes son las cuerdas)
con otra con constante de acoplamiento fuerte
(cuyos entes f\'isicos do\-mi\-nan\-tes son las branas).
Estas dualidades no cambian los radio de las dimensiones
compactificadas.
Por ejemplo, tenemos la teor\'ia tipo II-B consigo misma,
la heter\'otica $SO(32)$ con la tipo I y
la tipo II-A de acoplamiento fuerte con la
{\it \palabra{supergravedad} en once dimensiones} (SUGRA),
que es una formulaci\'on supersim\'etrica de GR.
A diferencia de las supercuerdas, once no representa la \'unica
dimensi\'on del espacio-tiempo para una formulaci\'on consistente
matem\'aticamente de SUGRA, pero si representa el n\'umero m\'aximo de
dimensiones para que sea posible bajo ciertos criterios t\'ecnicos.
En la secci\'on \ref{sec:branas}, mencionaremos c\'omo es posible la
dualidad a pesar de que las dimensiones de los espacio-tiempos no
coincidan.

\item[Dualidades U] {\ }

Son las que relacionan las teor\'ias de supercuerdas de cualquier
manera, sin importar que los radios y las constantes de acoplamientos
cambien.
Las dualidades T y S ser\'ian entonces casos particulares de este tipo
de dualidad.

\end{description}


\section{Teor\'ia M}

\begin{figure}[!htp]
\begin{center}
\includegraphics{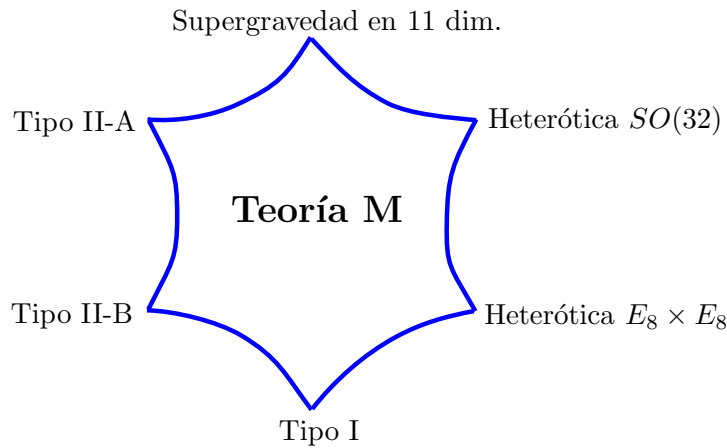}
\end{center}
\caption{Representaci\'on gr\'afica de la teor\'ia M.}
\label{fig:Mteoria}
\end{figure}

Como las \palabra{dualidades} sugieren que todas las teor\'ias de supercuerdas
son de alg\'un modo lo mismo, es entonces que proponemos a la
{\it \palabra{teor\'ia M}} como la ``mam\'a" de todas ellas.
El sentido de la propuesta es que la teor\'ia M es el nombre gen\'erico
para referirnos a todas las supercuerdas y estas son casos particulares
de ella, de acuerdo principalmente a constantes de acoplamiento, el
espacio de compactificaci\'on usado $X^6$\ldots\
Adem\'as de las cinco teor\'ias de supercuerdas, debemos incluir a
\index{supergravedad} SUGRA, porque es equivalente a la tipo II-A
seg\'un la dualidad S.
La Fig.~\ref{fig:Mteoria} muestra la manera popular de representar 
gr\'aficamente la teor\'ia M, como una entidad de la que conocemos
seis de sus ``puntas", de ah\'i la comparaci\'on con la historia de
``seis" ciegos que tocan un elefante e intentan describirlo:
la descripci\'on depender\'a de la parte que est\'en tocando,
uno de ellos lo definir\'a a trav\'es de las orejas, otro por la trompa,
el siguiente por los colmillos\ldots\
Al final, cada uno de ellos concluir\'a que hablan de algo diferente,
cuando realmente lo que han ``visto" son ciertas partes de un elefante.
De ese mismo modo es nuestro conocimiento de la teor\'ia M, le hemos
``visto" partes, pero todav\'ia no sabemos que es en su totalidad.
Por el comportamiento de las supercuerdas a nivel de acoplamiento
fuerte, concluimos que el espacio-tiempo de la teor\'ia M debe
ser de once dimensiones, pero su formulaci\'on exacta es todav\'ia un
problema abierto y activo dentro de la f\'isica.

Actualmente, el nombre de ``M" tiene muchas razones de ser, entre ellas
(por la palabra en ingl\'es):

\begin{itemize}

{\it \item Madre} (Mother) de todas las teor\'ias.

{\it \item Misterio} (Mystery), porque todav\'ia falta mucho por conocer
de la teor\'ia misma.

{\it \item ``W" invertida}, por Witten, el f\'isico te\'orico m\'as representativo
en teor\'ia de cuerdas.

{\it \item Matriz} (Matrix), porque es representable como una teor\'ia
cu\'antica de matrices, siendo la formulaci\'on m\'as prometedora
actualmente.

{\it \item Membrana} (Membrane), porque tiene otros entes extendidos,
adem\'as de las cuerdas.

\end{itemize}

Cabe mencionar que en cierta literatura de supercuerdas,
sobre todo en los primeros art\'iculos,
el nombre de teor\'ia M es a veces usado en referencia a SUGRA,
como un sin\'onimo en el sentido que a bajas energ\'ias
la teor\'ia M se describe cl\'asicamente usando SUGRA,
a diferencia del significado m\'as moderno de totalidad o entidad.
Una idea similar es el nombre de teor\'ia F para la que es
equivalente a la tipo II-B y que tiene once dimensiones,
recuerde que SUGRA (o teor\'ia M bajo lo comentado en este p\'arrafo)
es equivalente a la tipo II-A.


\section{Las \palabra{branas} \label{sec:branas}}

\begin{figure}[!htp]
\begin{center}
\includegraphics[width=5.4in]{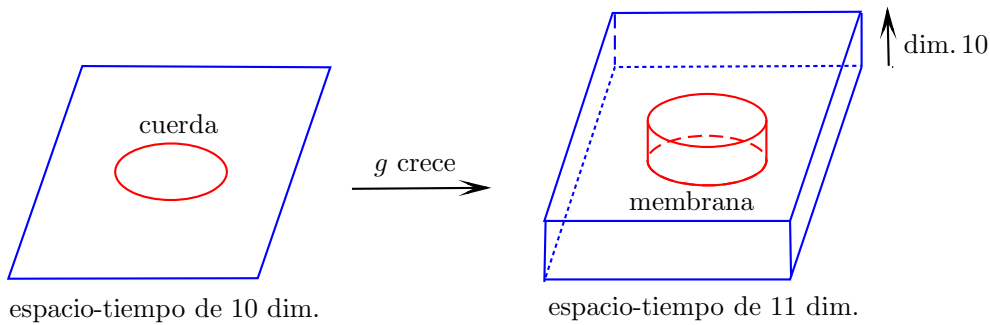}
\end{center}
\caption[Aparici\'on de una dimensi\'on extra por ``ensanchamiento".]{Aparici\'on
de una dimensi\'on extra por ``ensanchamiento" del
espacio-tiempo en la teor\'ia heter\'otica de $E_8 \times E_8$.}
\label{fig:branaE8}
\end{figure}

\begin{figure}[!htp]
\begin{center}
\includegraphics[width=5.4in]{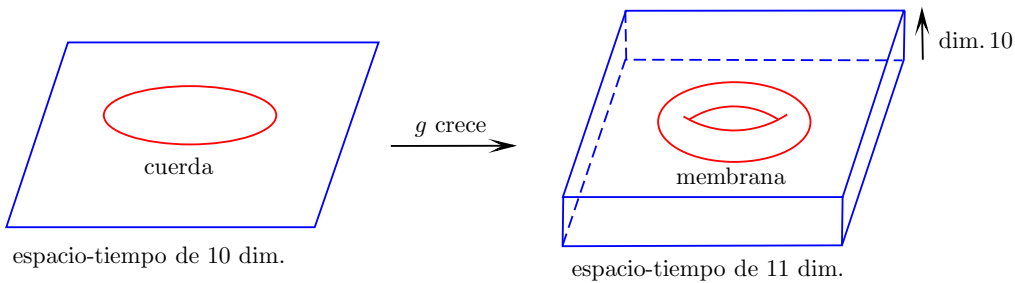}
\end{center}
\caption[Aparici\'on de una dimensi\'on extra por ``inflar".]{Aparici\'on
de una dimensi\'on extra por ``inflar" las
cuerdas en la teor\'ia tipo II-A.}
\label{fig:branaIIA}
\end{figure}

Expliquemos c\'omo es posible pasar de una teor\'ia de diez
dimensiones a otra de once dimensiones,
el artificio de hacer aparecer {\it una nueva dimensi\'on} est\'a en
la \palabra{constante de acoplamiento}, que llamaremos $g$.
Es posible mostrar que conforme hacemos $g$ m\'as grande
(acoplamiento fuerte),
una nueva dimensi\'on comienza a hacerse visible.
Las Figs.~\ref{fig:branaE8} y \ref{fig:branaIIA} muestran
gr\'aficamente dos modos de c\'omo una dimensi\'on extra puede aparecer.
En la Fig.~\ref{fig:branaE8}, el espacio-tiempo con todo lo que
contenga comienza a ganar ``ancho" conforme $g$ crece;
as\'i, por ejemplo, una cuerda cerrada pasar\'a a ser una membrana,
en este caso un cilindro.
La dimensi\'on extra la denotamos como dim.\,10.
En la Fig.~\ref{fig:branaIIA}, las cuerdas se ``inflan" al aumentar
$g$, por lo tanto, el espacio-tiempo est\'a obligado a incorporar otra
dimensi\'on para poder contener a las membranas, que en este caso
ser\'ian las cuerdas ``infladas" o donas
(correctamente, toros).
Este ``inflar" es el mecanismo presente en la teor\'ia tipo II-A que
le produce una dimensi\'on extra, nuestra imposici\'on anterior de que
la dimensi\'on de las supercuerdas era diez fue resultado de una
\palabra{formulaci\'on perturbativa}
(consider\'abamos que $g$ es peque\~no y hac\'iamos los c\'alculos de
los estados f\'isicos, las anomal\'ias\ldots\ como una serie en $g$),
de ah\'i nuestra imprecisi\'on.

Generalicemos ahora el concepto de membrana que ha sido usado para
cualquier superficie que tiene dos dimensiones,
como fueron anteriormente la parte externa del cilindro o del toro,
nuevamente no incluimos como parte de ellas lo que est\'e adentro,
en ese sentido ser\'ian como recipientes, son huecos.
Denominaremos entonces a la \index{branas} {\it brana} como
cualquier ente extendido de $p$ dimensiones espaciales
(recuerde que hemos incluido al tiempo como otra dimensi\'on,
por eso especificamos espaciales, no incluimos al tiempo para definirlas)
que adem\'as tiene una tensi\'on asociada,
o m\'as apropiadamente una densidad de energ\'ia.
En la formulaci\'on de una teor\'ia de cuerdas,
la tensi\'on desempe\~na un rol muy similar al de
la masa en las teor\'ias de part\'iculas puntuales.
Con esta nomenclatura, llamar\'iamos:
\mbox{\it 0-branas} a los puntos,
\mbox{\it 1-branas} a las cuerdas,
\mbox{\it 2-branas} a las membranas y
\mbox{\it p-branas} a los entes extendidos en $p$ dimensiones espaciales.

?`C\'omo explicamos que las branas no estuvieran desde el principio
en nuestra teor\'ia de supercuerdas? ?`No tuvimos el cuidado de
incorporarlas dentro de nuestra formulaci\'on?
Esta paradoja es descifrable porque las branas al ser entes f\'isicos
tienen masa, la cual es inversamente proporcional a la \palabra{constante de 
acoplamiento} de las cuerdas, es decir, a menor valor de $g$
(acoplamiento d\'ebil) mayor masa y viceversa.
Por tanto, como las teor\'ias de supercuerdas las formulamos de manera
perturbativa ($g$ es peque\~no),
la masa de las branas ser\'a demasiado grande,
lo que provoca un efecto insignificante en la f\'isica,
su ausencia es debido a que requieren de mucha energ\'ia
para ser creadas, contrario a las cuerdas.
Resumiendo, la \palabra{formulaci\'on perturbativa} de las supercuerdas no
define adecuadamente la teor\'ia, por tanto, la investigaci\'on
futura de las mismas har\'a que completemos todas las partes del
``elefante" llamado \palabra{teor\'ia M}.


\section{Interacci\'on entre cuerdas y branas \label{sec:cuerdabrana}}

\begin{figure}[!htp]
\begin{center}
\includegraphics{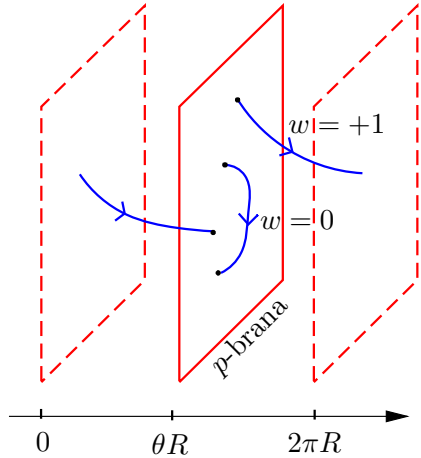}
\end{center}
\caption[Dos cuerdas abiertas atadas a una $p$-brana.]{Dos
cuerdas abiertas atadas a una $p$-brana.
Las cuerdas tienen n\'umeros de enrollamiento cero y uno.
Los planos de l\'ineas discontinuas est\'an identificados.
El eje representa la dimensi\'on compactificada.}
\label{fig:Dbrana}
\end{figure}

\begin{figure}[!htp]
\begin{center}
\includegraphics[height=2.2in]{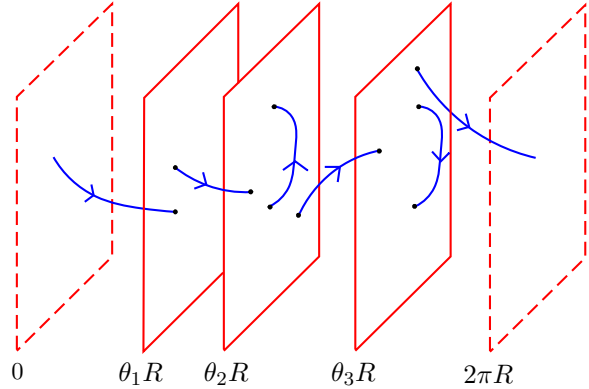}
\end{center}
\caption[Cinco cuerdas abiertas atadas a tres $p$-branas.]{Cinco
cuerdas abiertas atadas (interactuando) a tres $p$-branas.
Todas las cuerdas tienen $w = 0$.
Los planos de l\'ineas discontinuas est\'an identificados y
las posiciones de las branas en la dimensi\'on compactificada
son las indicadas.
Si las puntas de la cuerda est\'an en la misma brana, la cuerda no
tienen masa, caso contrario s\'i tiene, debido a la tensi\'on
producida por las branas al estirarla.}
\label{fig:TresBranas}
\end{figure}

Una justificaci\'on menos t\'ecnica de porqu\'e deben haber \palabra{branas}
en una teor\'ia de cuerdas es que las puntas de una cuerda abierta
podr\'ian estar restringidas a moverse solo en cierta regi\'on del
espacio.
Y si consideramos que estas regiones pueden ser entes f\'isicos:
experimentan interacciones entre ellas mismas y con las cuerdas,
se mueven, se deforman\dots, concluir\'iamos que deben ser
ingredientes de la teor\'ia.
Basado en el estudio de las branas, es que descubrimos las dualidades
y cierto comportamiento de las supercuerdas a nivel de acoplamiento
fuerte.
Incluso si las puntas se mueven en todo el espacio, diremos
que la regi\'on es una 9-brana, la que llena todo el espacio.
Bajo esta premisa, deducimos que las cuerdas abiertas y las branas
deben interactuar como lo muestran las Figs.~\ref{fig:Dbrana}
y \ref{fig:TresBranas}.
Ahora nuestra representaci\'on del espacio con algunas dimensiones
compactificadas
(note que ahora no incluimos al tiempo)
es a trav\'es del concepto de {\it \palabra{identificaci\'on}}.
Por ejemplo, un cilindro se representar\'ia como un plano o una hoja de papel,
uno tendr\'ia que pegarle los bordes para formarlo,
por lo que decimos que sus bordes est\'an identificados:
si uno caminara en el plano hasta el borde derecho y
contin\'ua caminando, uno aparecer\'a en el borde izquierdo
a la misma altura (al estilo de los juegos de video),
en alusi\'on al dar vueltas en un cilindro.

\begin{figure}[!htp]
\begin{center}
\includegraphics[height=2.2in]{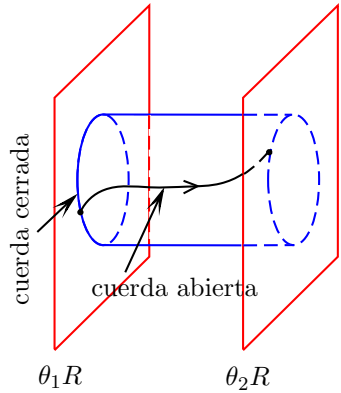}
\end{center}
\caption[Intercambio de una cuerda cerrada entre dos branas o
la correcci\'on cu\'antica para dos branas por un lazo de una cuerda
abierta.]{Intercambio de una cuerda cerrada entre dos branas o
la co\-rre\-cci\'on cu\'antica para dos branas por un lazo de una cuerda
abierta.
Las posiciones de las branas en la dimensi\'on compactificada son las
indicadas.
La cuerda cerrada traza un cilindro al ir de una brana a otra.
Cada punta de la cuerda abierta al moverse traza un c\'irculo en la
brana donde est\'a atada y su movimiento completo traza el cilindro.}
\label{fig:cerradabrana}
\end{figure}

Por otro lado, la Fig.~\ref{fig:cerradabrana} muestra el caso entre
cuerdas cerradas y branas, donde una cuerda cerrada (circular) se
intercambia de una brana a otra, es el \index{diagramas de Feynman}
diagrama de Feynman de la interacci\'on entre branas donde la
\index{part\'iculas mediadoras} mediadora es una cuerda cerrada;
lo que podr\'ia representar la interacci\'on gravitacional entre branas,
pues el gravit\'on se modela como cuerda cerrada.
Sin embargo, podemos tambi\'en interpretar el diagrama como el de
una cuerda abierta con sus puntas atadas a dos branas diferentes y
que al moverse forma un cilindro.
Este tipo de diagrama, llamado {\it de lazo} ({\it loop diagram}),
representa las correcciones cu\'anticas de un proceso;
en nuestro caso particular, la brana
se da cuenta de la presencia de la otra.
Esta doble interpretaci\'on es nuevamente resultado de las
equivalencias (topol\'ogicas) entre los diagramas, como lo mencionamos
al final de la secci\'on \ref{sec:string} y conlleva a confirmar los
c\'alculos pues podemos hacerlos por m\'etodos diferentes, o
deducirlos a trav\'es de un diagrama equivalente.

La consecuencia m\'as importante en la interacci\'on cuerdas-branas
es que nos permite incorporar apropiadamente el
modelo est\'andar de las part\'iculas
(ver introducci\'on) en una teor\'ia de cuerdas,
porque las QFT se pueden vincular directamente con el
{\it volumen de mundo} de las branas.
Estrictamente, a este tipo de brana se le denomina {\it D-brana},
porque su naturaleza est\'a asociada a las cuerdas abiertas
originales en la formulaci\'on de la teor\'ia.
Matem\'aticamente, son el resultado de las condiciones de frontera
de Dirichlet (por eso la letra D en su nombre).
Bajo este contexto, se han propuesto modelos te\'oricos
donde nuestro universo se considera como una D3-brana,
i.e.~somos seres que estamos confinados a vivir en esta
brana, el espacio de tres dimensiones que conocemos,
aunque realmente el universo tendr\'ia m\'as dimensiones espaciales.
Este tipo de enfoque es llamado
{\it brane-world approach} o {\it scenario}.


\section{Conclusiones}

Las supercuerdas son en la actualidad la mejor propuesta te\'orica
para unificar las teor\'ias cu\'anticas de campo (QFT) con la relatividad
general (GR), los dos grandes logros del siglo veinte en f\'isica, que
permiten explicar lo peque\~no (part\'iculas, \'atomos\dots) y
lo grande (el universo, las galaxias\dots), respectivamente.
Un marco unificado nos dar\'ia la capacidad de comprender adecuadamente
al universo mismo, su evoluci\'on y todo lo que est\'a en \'el.

Nuestro conocimiento actual de las supercuerdas lo fundamentamos
principalmente en su formulaci\'on perturbativa, las cuerdas
interact\'uan (se aco\-plan) d\'ebilmente.
Bajo esta formulaci\'on, la teor\'ia de supercuerdas requiere que
el espacio-tiempo sea de diez dimensiones para evitar el problema
de anomal\'ias.
Sin embargo, y a pesar de imponer otros requisitos a la teor\'ia,
encontramos cinco versiones de supercuerdas que son matem\'aticamente
consistentes.

El concepto de compactificaci\'on que permite interpretar las
dimensiones extras ha sido medular en el descubrimiento de las
dualidades y las branas,
los nuevos entes f\'isicos m\'as all\'a de puntos y cuerdas.
Con las dualidades, las versiones de las supercuerdas pierden su
car\'acter individual e incluso llegan tambi\'en a incorporar a una
teor\'ia que originalmente no es de cuerdas, SUGRA.
Ahora todas estas teor\'ias quedan unificadas bajo un solo marco
te\'orico: la teor\'ia M.

Formular adecuadamente la teor\'ia M es el nuevo objetivo, pero
tenemos el inconveniente que no sabemos con exactitud c\'omo lograrlo.
Primero que todo no es estrictamente una teor\'ia de cuerdas, donde
obviamente sus entes f\'isicos esenciales son las cuerdas, pues
contiene 2-branas
(las cuerdas ``infladas" de la Fig.~\ref{fig:branaIIA})
y 5-branas como parte esencial.
Otras caracter\'isticas de la teor\'ia son que el espacio-tiempo
debe ser de once dimensiones y que al considerar ciertos l\'imites
de sus par\'ametros como $g$, los radios de $X^6$
(estrictamente de $X^7$) entre otros,
deben reducirla a una teor\'ia de supercuerdas.
Mientras tanto, la propuesta matricial parece (y puede) ser el marco
te\'orico correcto para definir la teor\'ia M.

La debilidad m\'as fuerte del concepto de cuerdas es su falta
de verificaci\'on experimental, tanto directa como indirecta.
En otras palabras, no existe en la actualidad experimento o fen\'omeno
f\'isico conocido que requiera fuertemente una explicaci\'on
``cuerdista", basta usar QFT o GR para entender la f\'isica del
problema.
La unificaci\'on de QFT y GR, que es el objetivo de cuerdas, es un
problema completamente te\'orico debido a que estas teor\'ias son
incompatibles entre ellas.
De ah\'i que utilizar las teor\'ias de cuerdas para entender mejor el
modelo est\'andar o la cosmolog\'ia
(dominadas por QFT y GR, respectivamente)
representa otro camino, activo en la f\'isica de hoy,
para guiar nuestro esfuerzo a una formulaci\'on m\'as adecuada
f\'isicamente de las supercuerdas y de la teor\'ia M,
basada en informaci\'on experimental y no matem\'atica, el camino 
tradicional y actual.
El problema de esta formulaci\'on tradicional es que las
matem\'aticas requeridas y sus aplicaciones tambi\'en ``sufren" de
falta de conocimiento apropiado y completo,
pues necesitamos conjuntamente varias de sus \'areas
(como topolog\'ia, geometr\'ia diferencial y teor\'ia de grupos, entre otras)
para establecerlas,
y con ello lograr un planteo formal y exacto de los c\'alculos.

Obviamente, ciertos detalles de las cuerdas son m\'as complicados y
complejos de c\'omo los hemos presentando, por lo que requieren de una
mayor profundizaci\'on para un lector con inter\'es de involucrarse
en el tema (ver las referencias).\footnote{%
Las referencias \cite{Greene:1999kj} -- \cite{Woit:2006js}
son libros de divulgaci\'on cient\'ifica,
donde \cite{Greene:1999kj} es el m\'as conveniente para
p\'ublico en general, algunos son de temas relacionados con
teor\'ia de cuerdas como \cite{Susskind:2005js} y
tambi\'en se han incluido \cite{Smolin:2006pe} y \cite{Woit:2006js}
para contrastar con el otro punto de vista:
f\'isicos que no apoyan las cuerdas como una teor\'ia
con significado real para la f\'isica.
Los libros \cite{McMahon:2009zza} y \cite{Zwiebach:2004tj}
son de nivel de grado (bachillerato),
adecuados para iniciarse en el tema de las cuerdas,
donde personalmente se recomienda el segundo de ellos.
Los siguientes \cite{Becker:2007zj} -- \cite{Tong:2009np}
son de nivel de posgrado,
donde se recomienda \cite{Becker:2007zj} por tener
una estructura y presentaci\'on m\'as moderna de las cuerdas,
sin omitir los tradicionales \cite{Green:1987sp} y \cite{Polchinski:1998rq},
algunos son versiones que se pueden obtener libremente
en el internet.
Por otro lado, las referencias \cite{Hashimoto:2012vsa} -- \cite{Szabo:2004uy}
se enfocan fuertemente en el estudio de las branas y
sus aplicaciones dentro de la f\'isica,
principalmente en f\'isica de part\'iculas y cosmolog\'ia;
se recomienda \cite{Johnson:2003gi}.
Los art\'iculos \cite{Marolf:2003tx} -- \cite{Susskind:2003kw}
presentan rumbos que ha tomado la investigaci\'on en cuerdas,
donde \cite{Marolf:2003tx} es una buena gu\'ia relacionada
a la literatura principal de cuerdas hasta el 2004,
incluye p\'aginas web.
Finalmente, las \'ultimas dos referencias son ejemplos de cursos:
\cite{STfU:ZG} a nivel de grado (MIT) y
\cite{GCiST:U} a nivel de posgrado (Universidad Aut\'onoma de Madrid).
En ambos, el lector encontrar\'a los materiales correspondientes
y apreciar\'a la estructura de los cursos.
Cabe mencionar que estas referencias no pretenden ser la versi\'on
completa de la literatura m\'as importante y esencial de cuerdas.
}
Consecuentemente, este art\'iculo pretende ser solo
un punto de partida al mundo de las cuerdas,
cuya iniciativa inici\'o en los cursos de T\'opicos de F\'isica Te\'orica
que se imparten en la Universidad de Costa Rica.


\end{document}